\documentclass[11pt]{article}
\usepackage{latexsym}
\usepackage[dvips]{graphicx}
\usepackage[english]{babel}
\usepackage{epsfig}
\begin{document}
\title{Predictions of the Generalized Glauber Model for the
coherent $\rho$-production at RHIC and the STAR data
}
\author{
L. Frankfurt\\
\it School of Physics and Astronomy, Raymond and Beverly Sackler\\
\it Faculty of Exact Science, Tel Aviv University, Ramat Aviv 69978,\\
\it Tel Aviv , Israel\\
M. Strikman\\
\it Pennsylvania State University, University Park, Pennsylvania 16802,
\it USA\\
M. Zhalov\\
\it Petersburg Nuclear Physics Institute, Gatchina 188350, Russia}
\date{}
\maketitle

\centerline {\bf ABSTRACT}

{We calculate the rapidity distribution and the total cross 
section of coherent and incoherent  $\rho$-production
in the  heavy ion ultraperipheral collisions (UPC) at 
$\surd s_{NN}=130$ GeV using the generalized vector dominance 
model(GVDM) and the  
Gribov-Glauber
approach.  We find  the coherent 
cross section of $\rho $-production $\sigma_{coh}=490$ mb
compared to the $\sigma_{coh}=370\pm 170\pm 80$ mb recently reported by
the STAR collaboration at RHIC. 
The predicted cross section inside  the acceptance of the experiment,
 $\left | y\right |\le 1$, agrees with the data
within half a standard deviation.
It is emphasized that  measurements of
the rapidity distribution will provide a much more stringent
test of the model.}

\section{Introduction}
Ultraperipheral collisions(UPC) of relativistic heavy ions at RHIC 
and LHC open a promising new avenue for experimental studies 
of the photon induced coherent and incoherent interactions with nuclei
at high  energies (see \cite{baur,Baur2002,FELIX} for 
the reviews and extensive lists of references). Really, the  LHC heavy ion
program will allow studies of photon-proton and  photon - nucleus collisions 
at the energies exceeding by far those available now at HERA for $\gamma-p$ 
scattering.

Hence, it is very important to check our basic understanding of the UPC 
processes using the reactions which have smaller theoretical uncertainties on
the level of the $\gamma A$ interactions. Recently the STAR
collaboration released the first data on 
the cross section of the coherent $\rho $-meson production in gold-gold 
UPC at $\surd s_{NN}=130$ GeV \cite{star},\cite{star1}. 
This provides a first opportunity to check the basic features 
of the theoretical models and main approximations  which include the 
Weizs\"{a}cker-Williams
(WW) approximation for the spectrum of the equivalent photons, an 
approximate  procedure for removing collisions at small impact parameters
where nuclei interact strongly, and the model for the vector meson production
in the $\gamma A $ interactions. In the case of  the $\rho$-meson production
the basic process is understood much better than for
other photoproduction processes. Hence, checking the theory
for this case is especially important for 
proving that UPC could be used for learning new information 
about photon - nucleus interactions.

Earlier we published \cite{FSZrho}  predictions  for the cross section
of this process  at higher energy $\surd s_{NN}=200$ GeV.
Hence, a direct comparison of the STAR result with \cite{FSZrho}  is 
difficult. 
In this paper we perform the analysis  of the $\rho$-production at
$\surd s_{NN}=130$ GeV including the effects due to the  cuts
of the STAR experiment.

\section{Outline of the model}
 
 Production of $\rho $-mesons in ultraperipheral heavy ion collisions 
can be expressed in the Weizs\"{a}cker-Williams approximation\cite{ww}
through the $\gamma A\rightarrow \rho A$ cross section
\begin{eqnarray}
{\frac {d\sigma_{A_{1}A_{2}\to A_{1}A_{2}V}} {dy}}=
n_{A_{1}}^{\gamma }(y)\sigma_{\gamma A_{2}\to \rho A_{2}}(y)+
n_{A_{2}}^{\gamma }(-y)\sigma_{\gamma A_{1}\to \rho A_{1}}(-y).
\end{eqnarray}
The quantity $y=ln \frac {2\omega_{\gamma } } {M_{\rho }}$ is the rapidity 
of the produced $\rho $ meson and $n^{\gamma }(y)$ is the flux of photons 
with the energy $\omega _{\gamma }=\gamma _{c}q_0$ 
emitted by one of nuclei ($\gamma _{c}$ is 
the Lorentz factor for colliding nuclei,
and
$q_{0}$ is the photon momentum in the coordinate system of moving nucleus). 
The photoproduction cross section $\sigma_{\gamma A\to \rho A}(y)$ can be 
calculated in the Glauber model \cite{glauber}
\begin{eqnarray}
\sigma_{\gamma A\to \rho A}(y)  =\int \limits_{-\infty}^{t_{min}} dt
\frac {\pi} {k_{\rho}^2}{\left |F_{\gamma A\to \rho A}(t)\right |}^2=\frac {\pi} {k_{\rho}^2}
\int \limits_{0}^{\infty} dt_{\bot }
{\left|\frac {ik_{\rho}} {2\pi}  \int \,d\,{\vec b}\,  e^{i{\vec q}_{\bot }\cdot {\vec b}}
\Gamma ({\vec b}) 
\right|}^2.
\label{crosec}
\end{eqnarray}
Here 
${\vec q}_{\bot }^2=t_{\bot }={t_{min}-t}$, $-t_{min}=\frac {M_{\rho }^4} {4q_0^2}$ 
is longitudinal momentum transfer in $\gamma -\rho $ transition, 
and $\Gamma ({\vec b})$ is the diffractive nuclear profile function
\begin{eqnarray}
\Gamma ({\vec b})=\lim _{z\rightarrow \infty}\Phi ({\vec b},z).
\label{gamma}
\end{eqnarray}  
To calculate the eikonal function $\Phi ({\vec b},z)$ 
the Glauber approach 
\cite{bochmann} 
was  combined with the generalized vector dominance (GVD) model \cite{Gribov}.
More properly such an approximation should be called the Gribov-Glauber 
model\cite{Gribovinel} because the space-time evolution of high energy
processes is different in 
quantum mechanical models and in   quantum field
theory and therefore 
theoretical foundations for the high-energy  model are different.
In particular, in QCD in difference from quantum mechanics a high-energy 
projectile interacts with all nucleons at the same impact parameter 
almost at the same time \cite{Gribovinel}.  Such formulae allow to 
extend the domain of applicability of  the 
Glauber model to the description of
high energy phenomena where inelastic (high multiplicity) 
particle production gives dominant contribution
to  the total cross section. 
As we are 
mostly  interested in the accurate 
estimate
of the coherent diffractive production of the vector meson 
$\rho $ with $M_{\rho }=0.77$ GeV,  the spectrum of the higher corresponding 
hadronic states of $M\leq 2$ GeV can be approximated by one effective 
$\rho ^\prime$-meson with some reasonable fixed mass, say 
$M_{\rho ^\prime}=\sqrt {3}M_{\rho}$\cite{Pautz:qm}.
We want to draw attention that the value and sign of the  $\rho'$ 
contribution as taken from quenched GVDM is fitted to
describe  approximate Bjorken scaling in the case of highly 
virtual photon. Thus the model used in the paper correctly accounts
for the fluctuations of cross section including color transparency
phenomenon \cite{FGS}.
    
Then the GVD model comprises elementary  amplitudes 
\begin{eqnarray}
f_{\gamma N \to \rho N}=\frac {e} {f_{\rho }} f_{\rho N\to \rho N}+
\frac {e} {f_{\rho ^\prime}}f_{\rho ^\prime N\to \rho N}, 
\nonumber
\\
f_{\gamma N\to \rho ^\prime N}=
\frac {e} {f_{\rho ^\prime}} f_{\rho ^\prime N\to \rho ^\prime N}+
\frac {e} {f_{\rho }}f_{\rho N\to \rho ^\prime N}.
\label{gvdm}
\end{eqnarray}
In the optical limit ($A\gg 1$), 
with accuracy $O(\surd \alpha_{em})$ the eikonal functions 
$\Phi_{\rho ,\rho ^\prime} ({\vec b},z)$ 
are determined by the solutions of the coupled two-channel  equations  
\begin{eqnarray}
2ik_{\rho } \frac {d} {dz} \Phi _{\rho }({\vec b},z)=
U_{\gamma A\to \rho A}({\vec b},z)
e^{iq_{\| }^{\gamma \rightarrow \rho }z}+
\nonumber
\\ 
+U_{\rho A\to \rho A}({\vec b},z)\Phi _{\rho }({\vec b},z)+ 
U_{\rho A\to \rho \prime A}({\vec b},z)
e^{iq_{\|}^{\rho \rightarrow \rho \prime}z}\Phi _{\rho \prime}({\vec b},z)
\label{eqrho}
\end{eqnarray}
\begin{eqnarray}
2ik_{\rho \prime} \frac {d} {dz} \Phi _{\rho \prime }({\vec b},z)=
U_{\gamma A\to \rho \prime A}({\vec b},z)
e^{iq_{\| }^{\gamma \rightarrow \rho \prime}z}+
\nonumber
\\ 
+U_{\rho \prime A\to \rho \prime A}({\vec b},z)\Phi _{\rho \prime}({\vec b},z)+ 
U_{\rho \prime A\to \rho A}({\vec b},z)
e^{iq_{\| }^{\rho \prime \rightarrow \rho }z}\Phi _{\rho }({\vec b},z)
\label{eqrhop} 
\end{eqnarray}
with the initial condition $\Phi _{\rho ,\rho ^\prime} (\vec b ,-\infty)=0$.
The exponential factors $exp [iq_{\| }^{i\rightarrow j}z]$
are responsible for the coherent length effect, $i,j=\gamma ,\rho ,\rho ^\prime$, 
 $q_{\| }^{i\rightarrow j}=\frac {M_j^2-M_i^2} {2\gamma _{c}\omega }$.
The generalized Glauber 
-based optical potentials in the short-range approximation
are given by the expression
\begin{eqnarray}
U_{iA\to jA}({\vec b},z)=-4\pi f_{iN\to jN}(0)\varrho ({\vec b},z).
\end{eqnarray}
Here $f_{iN\to jN}(0)$ are the forward
elementary amplitudes, and $\varrho (\vec b,z) $ 
is the nuclear density normalized by the 
condition $\int d\vec b dz\,\varrho (\vec b,z)=A$.
We calculated $\varrho (\vec b,z)$ in the Hartree-Fock-Skyrme  (HFS) 
model which provided a very good(with an accuracy $\approx 2\%$) description 
of the global nuclear properties of spherical nuclei along the periodical 
table from carbon to uranium\cite{HFS} and the shell momentum distributions 
in the high energy (p,2p)\cite{p2p} and (e,e'p)\cite{eep} reactions. 

Following the simple suggestion of Ref.\cite{Pautz:qm} which is quite
reasonable in the  case of light vector mesons, we fixed  
the elementary scattering amplitudes and coupling constants by relations
\begin{eqnarray}
 f_{\rho ^\prime N\to \rho ^\prime N}=f_{\rho N\to \rho N},\qquad
f_{\rho N \to \rho^\prime N}=f_{\rho ^\prime N\to \rho N}=-\varepsilon f_{\rho N\to \rho N},
\qquad f_{\rho^{\prime}}=\frac {M_{\rho ^\prime}} {M_{\rho }} \cdot f_{\rho},
\label{mix}
\end{eqnarray}
with the value of the $\rho $-meson coupling constant
${f^2_{\rho}}/{4\pi}$=2.01. 
The diagonal amplitude $f_{\rho N\to \rho N}$ was taken in parameterization of
the Landshoff-Donnachie model\cite{LD} while the value of parameter 
$\varepsilon=0.18$
was found from a best fit to the differential cross section of
the $\rho$-meson photoproduction off lead at $\omega_{\gamma}=6.2$ GeV and
$t_{\bot }=0.001$ $GeV^2$\cite{MIT}. With all parameters fixed, we  
compared our 
calculations with all available data on $\rho $-meson photoproduction 
off nuclei 
at low and intermediate energies \cite{MIT} and found a very good description 
of the absolute cross section and of the momentum transfer distributions
(see \cite{FSZrho}). 
Hence, it was natural to expect that this model should   provide a reliable 
parameter-free predictions for production of $\rho $-mesons in high energy 
heavy ion UPC. Note here that the inelastic shadowing effects which 
start  to contribute at high energies still remain a few percent  
correction at energies $\le$ 100 GeV relevant for the STAR kinematics.
For LHC energy range one should account for the blackening of
interaction with nuclei. In this case cross section of inelastic 
diffraction in hadron-nucleus collisions should tend to 0. 
Also it leads to a suppression of the
$\rho'$ contribution to the  cross section of the  
diffractive $\rho$ meson photoproduction \cite{FSZrho}.

\section{Results and discussion}
\begin{figure}
    \begin{center}
        \leavevmode
        \epsfxsize=1.1\hsize
        \epsfbox{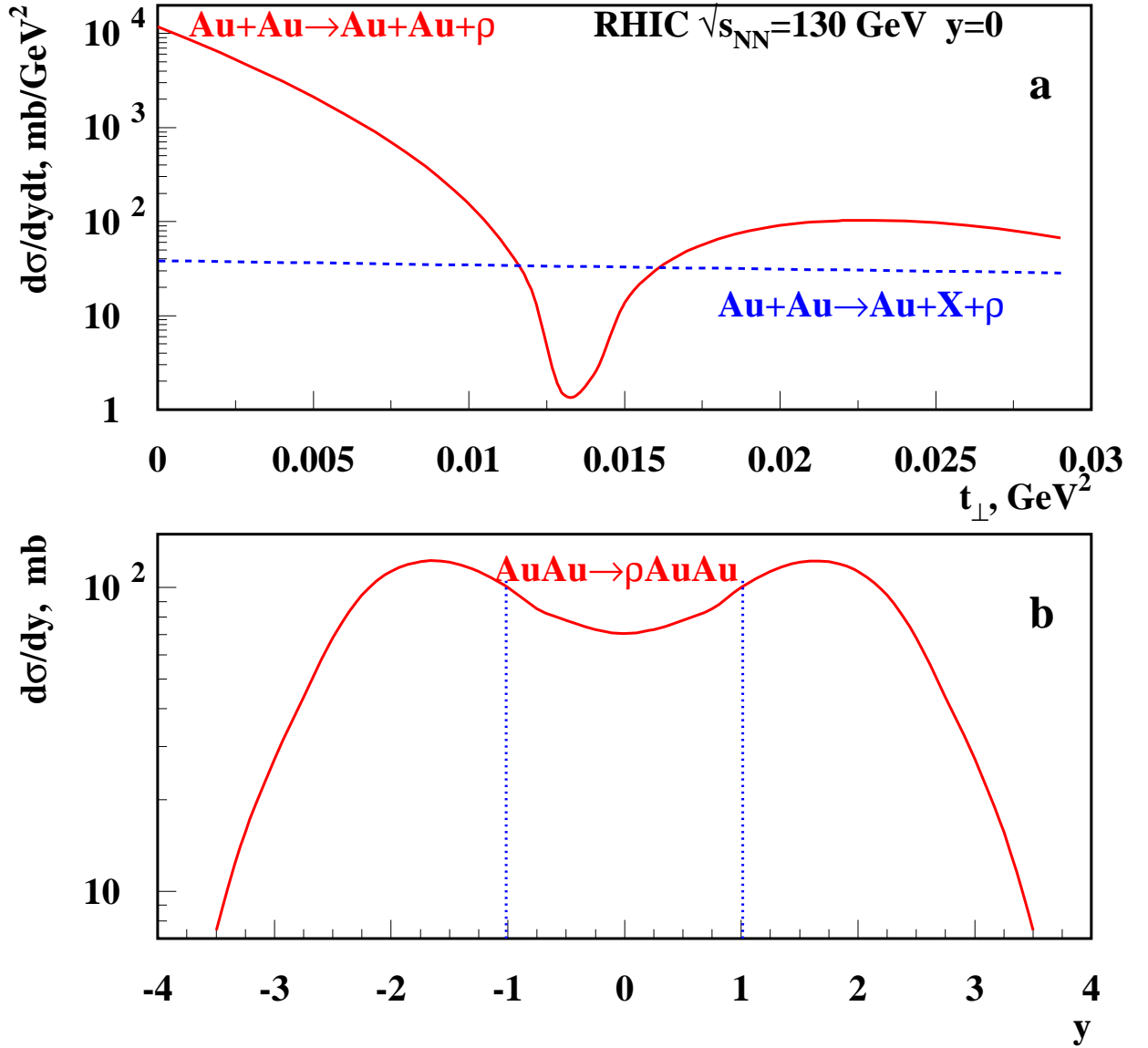}
    \end{center}
\caption{(a) Momentum transfer dependence of the coherent and
incoherent $\rho$-meson
production in AuAu UPC at $\surd s_{NN}=130$ GeV calculated in Generalized
Glauber model (GGM).
(b) Rapidity distributions for coherent
 $\rho $-meson production in the gold-gold UPC at $\surd s_{NN}=130$ GeV 
calculated in GGM.}
\label{dsdt}
\end{figure}

The  calculated momentum transfer distributions at the rapidity $y=0$ and 
the momentum transfer integrated rapidity distribution for gold-gold UPC 
at $\surd s_{NN}=130$ GeV  are presented in Figs.\ref{dsdt}a,b. 
 
\begin{figure}
    \begin{center}
        \leavevmode
        \epsfxsize=1.\hsize
        \epsfbox{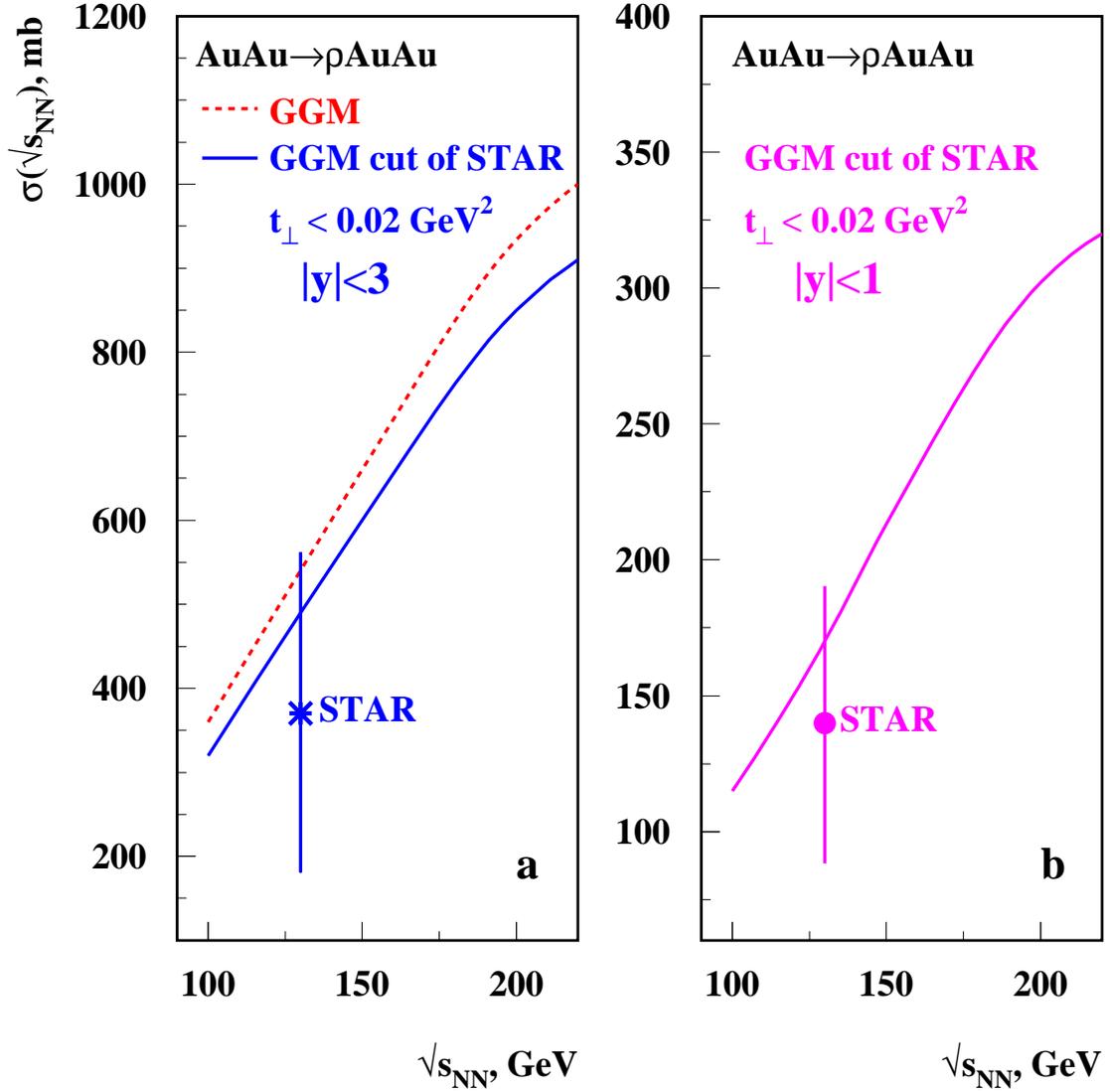}
    \end{center}
\caption{Energy dependence of the cross sections for coherent 
 $\rho $-meson production in the gold-gold UPC calculated in the GGM     
and the STAR results: a) The  dashed line is
the  total cross section in GGM, 
the solid      
line  is the  cross section calculated 
accounting for the STAR cut on the momentum     
transfer, star  is  the STAR cross section based 
on the Monte-Carlo extrapolation  
of the measured value to the full detector acceptance;  
b) Comparison of the GGM cross section in the interval of rapidities           
$\mid y \mid \leq 1$ with the value measured by the STAR. }   
\label{endep}
\end{figure}

Let us briefly comment on our estimate of the incoherent 
$\rho $-meson production 
cross section. The momentum transfer distribution(dashed line 
in Fig.\ref{dsdt}a) is practically flat in the discussed $t_{\bot }$ range.
The total incoherent cross section obtained by integration over 
the wide range of $t_{\bot }$ is $\sigma _{inc}=120$ mb.
To select the coherent production the
cut $t_{\bot }\leq 0.02$ GeV$^2$ was used  
in the data analysis \cite{star}. 
Correspondingly, the calculated incoherent cross  section for
this region of $t_{\bot }$ is $\sigma _{inc}=14$ mb. Our calculations 
of incoherent production which are based 
on accounting for only the single elementary diffractive collision  
obviously present the lower limit. The residual nucleus will be 
weakly excited and can evaporate only one-two neutrons.  
The events $A+A\to \rho+xn+A_{1}+A_{2}$  were detected 
by the STAR and identified as a 
two-stage process - coherent $\rho $-production
with the subsequent electromagnetic excitation and neutron 
decay of the  colliding nuclei \cite{Baltz2002}. 
In particular, the 
cross section estimated by the STAR for the  case when only one of 
the nuclei 
is excited and emits several  neutrons is 
$\sigma ^{\rho }_{xn,0n}=95\pm 60\pm 25$ mb.   
The momentum transfer distribution for these events is determined by
the  coherent production.  Hence, it differs from that
for incoherent events but in the region of very low  $t_{\bot }$
it is hardly possible  to separate 
them experimentally and obviously the measured cross section 
$\sigma ^{\rho }_{xn,0n}$
includes contribution
of incoherent events on the level of $15\%$.     

The total rapidity-integrated cross section of coherent $\rho $-meson
production calculated in the GVDM for the range of energies available at RHIC
is shown in Fig.\ref{endep}a (dashed  line).
We find $\sigma _{coh}^{th}=540$ mb at $\surd s_{NN}=130$ GeV. 
The  value $\sigma _{coh}^{exp}=370\pm 170\pm 80$ mb
was obtained at this energy by the STAR from 
the data analysis at the low momentum transfer $t_{\bot }\leq 0.02$ GeV$^2$. 
Thus, before making a comparison we should take into account
this cut.
It leads to a reduction of the  cross section  
by $\approx 10\%$ (the solid line in Fig.\ref{endep}a).
In our calculations we did not account for the $t_{\bot }$-dependence
of the elementary amplitudes which are rather flat in the considered
range of energies and momentum transfers 
as compared to that for the nucleus form factor.
So, in the region of integration important for our analysis
neglect by this slope is a reasonable approximation
but, nevertheless, an account of this effect would slightly 
reduce our estimate of the total cross section. Also we neglected a smearing 
due to the transverse momentum of photons and 
the interference of the production amplitudes from both nuclei 
\cite{Klein992}. 
This latter phenomenon results only in the narrow dip in the coherent 
$t_{\bot }$-distribution at 
$t_{\bot }\leq 5\cdot 10^{-4}$ GeV$^2$. All these effects
does not influence  noticeably  the value of the $t_{\bot }$-integrated
cross section but can be easy treated and taken into account in a more 
refined analysis. 
Thus we find $\sigma _{coh}^{th}=490$ mb to be compared to the STAR 
value $\sigma _{coh}^{exp}=370\pm 170\pm 80$ mb.
Since our calculation does not have any free parameters, this can be
considered as  a reasonable agreement.

At this point we would like to comment on the statement of \cite{star} that
our prediction for coherent $\rho $-production in gold-gold UPC at 
$\surd s_{NN}$=200 GeV\cite{FSZrho} is $50\%$ higher than the 
value given by model in\cite{Klein991}.
We already briefly explained in \cite{FSZrho}
that this   discrepancy originates from 
a number of approximations made  in
the model of Ref.\cite{Klein991} which differs from the 
conventional Glauber model.
The coherent photoproduction cross section was defined in\cite{Klein991}
by the expression:
\begin{eqnarray}
\sigma_{\gamma A\to \rho A}=
\frac {d\sigma_{\gamma A\to \rho A}(t=0)} {dt}\cdot
\int \limits_{-\infty}^{t_{min}}
{\left | F_{A}(t)\right |}^2dt.
\label{kleinmod}
\end{eqnarray}
where $F_{A}(t)$ is the 
nuclear form factor (two-dimensional Fourier transformation of the 
parameterized nuclear density)  and
the forward photoproduction cross section was estimated using
the vector dominance model and optical theorem 
\begin{eqnarray}
\frac {d\sigma_{\gamma A\to \rho A}(t=0)} {dt}=
\frac {\alpha _{em}} {4f^2_{\rho }}
\sigma^2_{tot}(\rho A).
\label{fcsklein}
\end{eqnarray}
The total cross 
section of the $\rho A$ interaction was found in\cite{Klein991} using 
the formula:
\begin{eqnarray} 
\sigma_{tot}(\rho A)=\int d{\vec b}\left [1-exp\biggl [-\sigma_{\rho N}
\int \limits_{-\infty}^{\infty} \varrho ({\vec b},z)dz \biggr ]\right ].
\label{cm}
\end{eqnarray}
Eq.(\ref{cm}) presents the  classical mechanics 
model with standard for this approach
expression for the
total cross section $\sigma^{cm}_{tot}(\rho A)$. The  
quantum mechanics expression  is given by the 
Gribov-Glauber model (here for simplicity we give the expression in 
the limit of  Re/Im=0)
\footnote
{In the Appendix we  
demonstrate how 
the model used in Ref.\cite{Klein991} but with the correct high energy 
expression for the total cross section can be obtained from the 
Gribov-Glauber based
Generalized
Vector Dominance Model and what essential approximations have to be
done on 
the way.}:  
\begin{eqnarray}
\sigma^{qm}_{tot}(\rho A)=2 \int d{\vec b}
\Biggl [1-exp\left[-\frac {1} {2} \sigma_{\rho N}T_{A}({\vec b})\right]\Biggr].
\label{qm}
\end{eqnarray}
In the Black Body Limit($\sigma_{VN}\to ~ \infty$) the total 
$\gamma A\to V A$ cross section estimated with the use 
of the classical mechanics ($\sigma^{cm}_{tot}=\pi R_{A}^2$) and the quantum
mechanics ($\sigma^{qm}_{tot}=2\pi R_{A}^2$) expressions 
in Eq.(\ref{fcsklein}) 
differ
by a factor of four. The difference for the  case of the gold nucleus
and reasonable value of the $\rho N$ elementary cross section 
$\sigma_{\rho N}\approx 25$ mb
can be found using the simplified model of the nucleus of constant density  
$\varrho _{0}\approx 0.16 \,fm^{-3}$ and radius $R_{A}$.
With the value of
 radius $R_{Au}\approx 6.5\,fm$ one can obtain
the reasonable estimate of the ratio
$$\frac {\sigma^{qm}_{tot}(\rho A)} {\sigma^{cm}_{tot}(\rho A)}
\approx 2\left [1-\frac {3} {2\sigma_{\rho N}^2 \varrho _{0}^2 R^2_{A}}\right ]\approx 1.55$$
Hence, we find that due to  
use of the classical mechanics expression  (\ref{cm}) instead of the 
Gribov-Glauber model
expression (\ref{qm})
the total $\gamma A\to \rho A$ cross section was underestimated in 
Ref.\cite{Klein991} by a factor $\approx 2.5$. Note, that in the range
of the photon energies essential in photoproduction of $\rho$-mesons in UPC 
at RHIC the elementary $\rho N$ cross section still weakly depends on
the energy of $\rho$-mesons. Hence, this factor weakly depends on
$\surd s_{NN}$. 
   
Next important approximation made in Ref.\cite{Klein991} 
in deriving 
formulae Eqs.(\ref{kleinmod}) and (\ref{fcsklein}) was the
 neglect by the  coherence 
length effects.
This requires $q_{\| }^{\gamma \to \rho}z=\frac {M_{\rho}^2z} {2{\gamma _{c}^2}q_0} \ll 1$.
This neglect is not justified because
large longitudinal distances are essential in the diffractive 
$\rho$-photoproduction. The $\rho$-meson can be formed far from the nucleus.
Besides, the photon flux is large at small $q_0\ll R_{A}^{-1}$, i.e. in the 
region where the coherence length effect is important. We estimated
that the cross section
for $AuAu\to \rho AuAu$ at the energy $\surd s_{NN}=200$ GeV
is overestimated
by a factor $\approx 1.5$  
if one neglects  the coherence length effect.  
The coherence length effect becomes  more essential 
with decrease of the energy.
As a result at $\surd s_{NN}=130$ GeV this effect suppresses the cross section
by a factor 
$\approx 2$. On the contrary,
 at much higher energies where
the coherence length is very large (for example, at LHC) 
this effect will be small.

   Note in passing
 that the calculations in Ref.\cite{Klein991} were performed 
neglecting  the real part of the elementary $\rho N$ amplitude which we
accounted for using the Landshoff-Donnachie parameterization. 
In the high energy
domain, for example, in the region of the central rapidities at RHIC
the real part of the $\rho N$ amplitude is negligible but one
 should account for
$Re f_{\rho N\to  \rho N}$   at the edges of rapidity
distribution which correspond to the 
photoproduction at intermediate energies of photons.
Since the contribution of this region to the total cross section is enhanced 
by the high photon flux, the total cross section of 
the coherent $\rho$ production 
at  $\surd s_{NN}=130$ GeV 
would be  underestimated by
$\approx 10\%$.

Thus we emphasize that a number of 
approximations were made
in the model of the Ref.\cite{Klein991} which are inconsistent with the 
conventional Glauber model and 
their interplay depends on the photon energy. Thus 
 the estimates obtained within this model
contain systematic uncertainties.
We want to draw attention that the
model we have used is well theoretically justified. It correctly 
calculates the nuclear form factor in the coherent photoproduction. 
We checked \cite{FSZrho}  that
this model provided a very good description of the
coherent $\rho $-photoproduction off nuclei at low and intermediate
energies along the periodical table
without any free parameters. 

At present the comparison  of our predictions
with STAR data is still preliminary because
the experimental errors are too large
and there are  few points in the procedure of the data analysis which 
should be discussed. 
Really the acceptance of STAR is very strongly $y$ dependent being
maximal at $y=0$
and going to zero at $\mid  y \mid  =1$ (Fig.3a in \cite{star})
while the theoretical distribution
is expected to have a double bump shape, see Fig.\ref{dsdt}b, which
is simply due to the symmetry of collision and  the interplay
of the energy dependence of the photon flux and $\gamma A\to \rho A$
cross section. 

Due to the acceptance conditions 
$AuAu\to \rho AuAu$ events were detected in the range of rapidities  
$|y|\leq 1$,  while the cross section reported in \cite{star}  is corrected 
for the $\mid y\mid \ge 1$
using the Monte-Carlo extrapolation based on the model \cite{Klein991}
in which 
\footnote{We thank S.Klein for the discussion of this issue.}
\begin{eqnarray}
R={\sigma_{4\pi }}/{\sigma_{\mid y \mid \leq 1}}=2.7.
\label{cut}
\end{eqnarray}
Thus the cross section in the region 
${\mid y \mid \leq 1}$ is
\begin{eqnarray} 
{\sigma_{\mid y \mid \leq 1}}=140\pm 60 \pm 30\,mb. 
\label{stcs}
\end{eqnarray}
The errors for this cut are scaled accordingly. 
However the model \cite{Klein991} differs from our in many aspects, 
described above. In particular, the
neglect by the coherence length effect 
leads to a significant modification of 
the value of factor $R$. 
The rapidity interval $\mid  y \mid \leq 1$
corresponds high  energies of the photon where this effect is negligible.
On the other hand due to 
the neglect by the coherent length effects 
 at the edges of the rapidity distribution the relative contribution of
photoproduction cross section 
at 
$\mid y \mid \geq 1$ 
was overestimated in Ref.\cite{Klein991}.
Hence their 
 value of $R$ is lower than the one  we obtain  
from our rapidity distribution  which gives $R=3.2$,
and the  cross section  
$\sigma_{\mid y \mid \leq 1}=170 $ mb.
It is this value of the cross section  which  should be compared 
(Fig.\ref{endep}b) 
to
the experimentally measured cross section (\ref{stcs}).
Obviously a more detailed comparison would require a  detailed study of
the sensitivity of the analysis to the assumed $y$-distribution.

\section{Conclusions}
We demonstrate that the cross section of coherent $\rho $-meson production
in high energy heavy ion UPC calculated within the GVDM  
is in a  good agreement
with experimental results of  the STAR collaboration. More stringent
test will involve comparison of the model predictions with the cross 
section and the rapidity distributions of the $\rho$-meson production 
measured with higher precision.

We thank A.~Baltz, B.~Kopeliovich, P.~Yepes,  and especially 
S.~Klein for very useful discussions. 
This study was supported in part by GIF, DOE and CRDF

\section{Appendix}
For the completeness let us discuss how classical mechanical formulae used in
Ref.\cite{Klein991} arise from the quantum mechanical ones.
Since 
the
Vector Dominance model was used in
\cite{Klein991} we should neglect 
by the nondiagonal $\rho \to \rho^\prime$ transitions 
in Eq.(\ref{gvdm}).
 Thus we need to put
$\varepsilon =0$ in Eq.(\ref{mix})
instead of $\varepsilon =0.18$ which 
is used in our calculations. If we keep all other parameters of the model fixed, 
we pay for such a reduction by the increase of the $\rho$-photoproduction 
cross section by a factor $\approx (1+2\varepsilon /\sqrt {3})\approx 1.2$. 
With $\varepsilon =0$ 
 equations (\ref{eqrho}) and (\ref{eqrhop}) 
become decoupled and the solution of Eq.(\ref{eqrho}) gives the
eikonal function of the $\rho $-meson
\begin{eqnarray}
\Phi _{\rho}({\vec b},z)=\frac {1} {2ik_{\rho}}e^{\frac {1} {2ik_{\rho}}
\int \limits_{-\infty}^{z}
U_{\rho N\to \rho N}({\vec b},z')dz'}
\int \limits_{-\infty}^{z}\,d\,z'
U_{\gamma N\to \rho N}({\vec b},z')e^{iq_{\| }^{\gamma \to \rho}z'}
e^{-\frac {1} {2ik_{\rho}}
\int \limits_{-\infty}^{z'}
U_{\rho N\to \rho N}({\vec b},z'')dz''}.
\label{frho}
\end{eqnarray}
Using 
 Eq.(\ref{crosec}) and Eq.(\ref{gamma}) and standard expression for the
elementary amplitude 
$$f_{\rho N}(0)=\frac {ik\sigma_{\rho N}} {4\pi}[1-i\beta_{\rho N}] \qquad
 \beta_{\rho N}=\frac {Re f_{\rho N}(0)} {Im f_{\rho N}(0)},$$
we obtain the amplitude of the $\rho$-meson  photoproduction  
off the nucleus in the optical limit of the standard Glauber plus 
the Vector Dominance 
model\cite{yenn}
\begin{eqnarray}
F_{\gamma A\to \rho A}=f_{\gamma N\to \rho N}(0)
\int \limits_{0}^{\infty} d\,{\vec b} e^{i{\vec q}_{\bot }\cdot {\vec b}}
\int \limits_{-\infty}^{\infty}dz' \,\varrho ({\vec b},z')
e^{iq_{\| }^{\gamma \to \rho}z'}e^{-
\frac {\sigma_{\rho N}} {2}[1-i\beta _{\rho N}]
\int \limits_{z'}^{\infty} \varrho ({\vec b},z'')dz''}.
\label{gampl} 
\end{eqnarray}
Note that accounting for the real part of the $\rho N$ amplitude leads to appearance
of the phase factor 
$exp[i\beta _{\rho N}\int \limits_{z}^{\infty} \varrho ({\vec b},z')dz']$ which is similar
to that describing the coherence length effect and 
which is important in the same energy domain. 
Following  the assumptions of Ref.\cite{Klein991} where both   
the coherence length effect and real part of the $\rho N$ amplitude were neglected
we remove these exponential factor from Eq.(\ref{gampl}).
 Then in
the limit of the purely imaginary elementary $\rho N$ amplitude 
 we obtain 
\begin{eqnarray}
F_{\gamma A\to \rho A}=f_{\gamma N\to \rho N}(0)
\int \limits_{0}^{\infty} d\,{\vec b} e^{i{\vec q}_{\bot }\cdot {\vec b}}
\int \limits_{-\infty}^{\infty}dz' \,\varrho ({\vec b},z')
e^{-\frac {\sigma_{\rho N}} {2}
\int \limits_{z'}^{\infty} \varrho ({\vec b},z'')dz''}=\\
\nonumber
\frac {f_{\gamma N\to \rho N}(0)} {\sigma_{\rho N}}\cdot 2 \cdot
\int \limits_{0}^{\infty} d\,{\vec b} e^{i{\vec q}_{\bot }\cdot {\vec b}}
\int \limits_{-\infty}^{\infty}dz \,\frac {d} {dz}
exp\biggl [{-\frac {\sigma_{\rho N}} {2}
\int \limits_{z}^{\infty} \varrho ({\vec b},z')dz'}\biggr ]=\\
=\frac {f_{\gamma N\to \rho N}(0)} {\sigma_{\rho N}}\cdot 2 \cdot
\int \limits_{0}^{\infty} d\,{\vec b} e^{i{\vec q}_{\bot }\cdot {\vec b}}
 \, \left [1-
exp\biggl [{-\frac {\sigma_{\rho N}} {2}
\int \limits_{-\infty}^{\infty} \varrho ({\vec b},z)dz}\biggr ]\right ]
.
\label{gvdam} 
\end{eqnarray}
Now using the vector dominance relation 
$f_{\gamma N\to \rho N}=\frac {e} {f_{\rho}^2}f_{\rho N\to \rho N}$ 
we can  write the formula for the forward $\gamma A\to \rho A$ cross 
section in the optical limit of the Glauber +VD model
\begin{eqnarray}
\frac {d\sigma_{\gamma A\to \rho A}(t=0)} {dt}=\frac {\pi} {k_{\rho}}
{\left |F_{\gamma A\to \rho A}(t=0)\right |}^2=
\frac {\alpha _{em}} {4f^2_{\rho }}\sigma_{\rho A}^2
\end{eqnarray}
where
\begin{eqnarray}
\sigma_{\rho A}=
2\cdot \int d\,{\vec b}\Biggl[1-
exp\biggl [{-\frac {\sigma_{\rho N}} {2}
\int \limits_{-\infty}^{\infty} \varrho ({\vec b},z)dz}\biggr ]\Biggr]
\label{gfwcs}
\end{eqnarray}
Thus in the Glauber+VD model we have got the expression for the 
forward cross section of the
photoproduction coinciding with Eq.(\ref{fcsklein}) used in calculations
performed in Ref.\cite{Klein991} but with 
high energy quantum mechanics formula for the total cross section.

\end{document}